# COSTS MODELS IN DESIGN AND MANUFACTURING OF SAND CASTING PRODUCTS

**Nicolas PERRY**
Ass. Prof., IRCCyN, 1 rue de la Noé, BP.92101, 44321 Nantes Cedex, France
33 (0)2 40 37 69 54, Nicolas.Perry@irccyn.ec-nantes.fr
**Magali MAUCHAND**
PhD Student, IRCCyN, 1 rue de la Noé, BP.92101, 44321 Nantes Cedex, France
**Alain BERNARD**
Professor, IRCCyN, 1 rue de la Noé, BP.92101, 44321 Nantes Cedex, France
33 (0)2 40 37 69 66, Alain.Bernard@irccyn.ec-nantes.fr

**Abstract:**

*In the early phases of the product life cycle, the costs controls became a major decision tool in the competitiveness of the companies due to the world competition. After defining the problems related to this control difficulties, we will present an approach using a concept of cost entity related to the design and realization activities of the product. We will try to apply this approach to the fields of the sand casting foundry. This work will highlight the enterprise modelling difficulties (limits of a global cost modelling) and some specifics limitations of the tool used for this development. Finally we will discuss on the limits of a generic approach.*

## 1   Introduction

In the early Seventies, studies in the United Kingdom and in the United States highlight the strategic role of the design activities. The conclusions lead both companies and authorities towards new approaches in order to improve the economic performances of companies. At the end of the Eighties, the paramount role of the quality in the design was reinforced in the United States by the Made-in-America report from the MIT "Commission on the Productivity". The Improving Engineering Design confirmed these conclusions in 1991: Designing for Competitive Advantage report, from the United State Nation Research Council "Engineering Design Theory and Methodology". As resumed by Perrin [1], the design phase is the key factor of the product development process. The ability to product new products with a high quality, a low cost and witch fit with the customer requests is fundamental to improve the nation competitiveness. Consequently, the costs (and cost management from the early design to the end delivery) become as important as the other technical requests.

Due to the global market and the worldwide competition, reactivity and agility are the only way to maintain the enterprise competitiveness. This can be characterized by the ability to change its products and/or processes in very short times and at minimal cost. The cost control, at the early step of design, becomes a key factor of success, since this phase fixes an average up to 70 to 80% of the end product costs (depending on the kind of production).

Moreover, the costs distribution (respectively direct and non direct) is changing: more time and services are dedicated to the studies for smaller products batches and shorter product life. The former fees sharing-out methods, the analytic or analogical cost-accounting methods, no longer give efficient results. Then, thanks to studies from CAM-I (Computer Aided Manufacturing-International) and authors like Johnson and Kaplan, the increasing gap between "traditional methods" of cost estimation and the new management requirements were highlighted.

All these works lead to new approaches integrating the complete cost and spread accounting methods based on the enterprise activities (ABC for instance). French economist, G.Perrin [2], since the Sixties, also developed a method based on one single cost-inductors through all the steps of the product development process (Added Value Unit method). We implemented such a costing management in a French sand casting foundry in order to allow a several-level management, based on indicators linked to the exact costs of the product to be delivered [3-4]. During this PHD thesis we validated the concepts but also the methodology required for a complete numerical traceability.

The work that will be presented in this paper concerns this former study and uses a concept called cost entity [5]. It includes several concepts: the cost inductors from the activity based accounting methods, the feature from the CAD and the homogeneity from the analytical cost accounting. Consequently, in order to define a cost entity, it is necessary to fill in several attributes linking technical and economical variables. The product model uses the concept of manufacturing feature. The cost are evaluated on the base of specific knowledge and reasoning models with the tool "*Cost Advantage*", giving information on costs to the CAD model. This is adding a cost semantic level to the CAD model. This models (called costgramme) put the expertise of the manufacturing cost available to the designer.

Some models, dedicated to the sand casting production of primary parts, were created with the wish to evaluate a meta-model that could be deployed in all the sand casting industries. Thus, the goals of this study are, on the one hand, to create the more generic as possible model related to the sand casting job, and on the other hand, to determine how they are transposable from a company to another (or from a production line to another…). So we will have to define and discuss which are the limits of the concepts from the triptych product/process/cost, and what level of detail is necessary to implement the methodology in most of the industrial environment.

This study is realized in collaboration with Cognition Europe, software developer of the tool *Cost Advantage*, and based on an industrial experimentation in the already-mentioned SMC Colombier Fontaine casting enterprise.

## 2   Value and cost management

The concept of cost or value strongly evolved with the context. Today, the selling price depends on the price the customer is ready to pay according to the value that it appreciates. Consequently, the margin is not used to calculate the selling price, it results from it. The selling margin and price must be defined beforehand; they determine an "objective maximum cost", known as "objective cost".

According to Perrin the indicators relating to the costs are mainly to be classified among the results' indicators [1]. The two criteria considered are on one side the cost of the product and on the other side the respect of the budget:

- the respect of the product cost is proposed within the framework of the development of the target costing methods (management by cost-target) and ABC (Activity Based Costing). The idea is to fix a target cost as an objective to which one must make correspond the cost of the product. The indicator then used, is the relationship between the effective and the objective product cost,

- the respect of the budgets, in particular of the studies' budgets, is the second element to be taken into account at the financial level. The indicator is then naturally the ratio between the exceeding and the initially fixed budget.

The contextual industrial evolutions change the nature of the costs in the company. Formerly, the final costs included a large majority of direct costs (often about 70%), i.e. directly assigned to the products (labour or raw material). The other costs could be the subject of global distributions; the choice of the scale hardly influenced the result. The method of the cost accounting with indirect expenses shared through a fixed percentage, can be admitted as long as their proportion is not too big. But the proportions are reversed, the direct costs do not constitute often any more that 30 % of the total cost. Not only the indirect expenses, according to odd sharing keys, create an arbitrary factor in the calculation of the costs, but it make the control difficult. These two concerns gave place to many researches that ended up in deeply modifying the systems of costs evaluation.

Let us analyse the product life cycle and the impact of decisions in term of generated costs. The early phase of design implies between 70 and 85% of the product cost and, at the end of the detailed design the margin on the final cost is thus very limited (cf. Fig 1). Contrary, the evolution of the costs really engaged by the company are very limited in these upstream phases. However, it is there that the control of the costs should be really exerted.

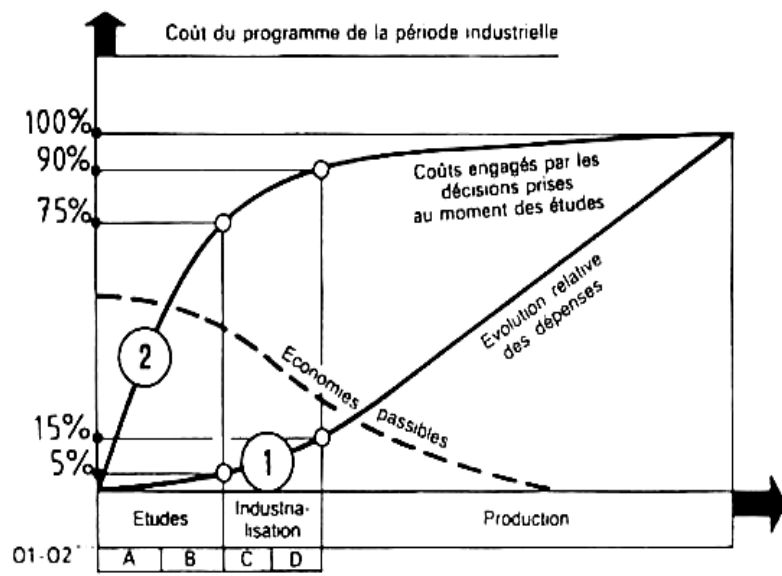

*Fig.1 : Cost evolution during the PLM [6]*

Consequently, the more the product is defined, the less the cost reduction is easy. The modification becomes increasingly expensive. There is thus an obvious economic interest to have a product optimisation at the early step of the process. The profits potentialities are significant and the committed costs weak. The design choices, inexpensive in terms of resources consumption can be very expensive in production, when it becomes extremely difficult to carry out modifications [7-8].

## 3   Design to Cost

The design to cost can be defined as a principle of action aiming the establishment of rigorous objectives. It allows compromises between performance and cost [9-10]. The "cost" constraint becomes capital and this data should be managed on the same level than technical performances. Then the cost objectives become constraints and the technical performances, variables.

The interest to integrate the cost management at the design phase is shared either by the customer and the seller since it makes it possible to control the development according to the exact needs for the future users (it incites in the search of new ideas required by the economic constraints), in addition, it makes possible to prepare and organize its production very early and to better control its margin [11].

This method of design takes again the steps of activities or tasks with a focus on the costs limits. This led to an extension of the design time due to the successive iterations needed to agree on the solutions after negotiation (technical or economic). This leads to cross the functions within the company and can be used jointly with methods of costs analysis like ABC. This also allows a very thorough analysis of the product functions linked to its costs [12].

One can however point out that the design phase is longer. But, the debugging step (pre-production) is no longer disturbed by modifications as it occurs in a traditional step. We then accept in fact "to waste" time with the design phase, therefore with the quasi-certainty to regain it downstream before the delivery date.

### 3.1 The Cost Entity concept and the modelling logic

The aim of our study is to sharply manage the costs (direct and indirect) during the production of sand casting parts. As illustrated previously, it is imperative to give a tool to the engineers of the engineering and design department in order to help them to control the costs of the parts design. In collaboration with the company Cognition Europe, and on the basis of the tool *Cost Advantage*, we work on the costs models to apply in the case of the steel sands casting parts. We are based on a preceding work, proposing an integrated approach for the sand foundry, realized within the framework of a thesis in partnership with SMC Colombier Fontaine (France) of group AFE Metal [2]. This work allowed formalizing the base of knowledge trade necessary to the control of the product life cycle in a foundry company. In addition, we validated an approach, a methodology and a deployment leading to ensure an exact knowledge of the parts costs and their impact on the output of the company [4].

Let us start with the concept of cost entity and context, which are our modelling bases.

#### 3.1.1   Costs Entities

A Cost Entity is a grouping of costs associated with the resources consumed by an activity. The general condition is due to the homogeneity of the resources, which makes it possible to associate a single inductor: the entity cost [5]. The model allows expertise formalisation, knowledge capitalisation and to have, at the early design phase, some information about the production step. Moreover, it helps the communication between all collaborators during the product life cycle.

### 3.1.2 Contexts

The contexts specify defined entities in three levels in our model. The first is defined in a process level, the second on a material level and the last is directly related to the feature. This context is a cross between a process, a material and a feature, connected to an environment (cf. Fig.2).

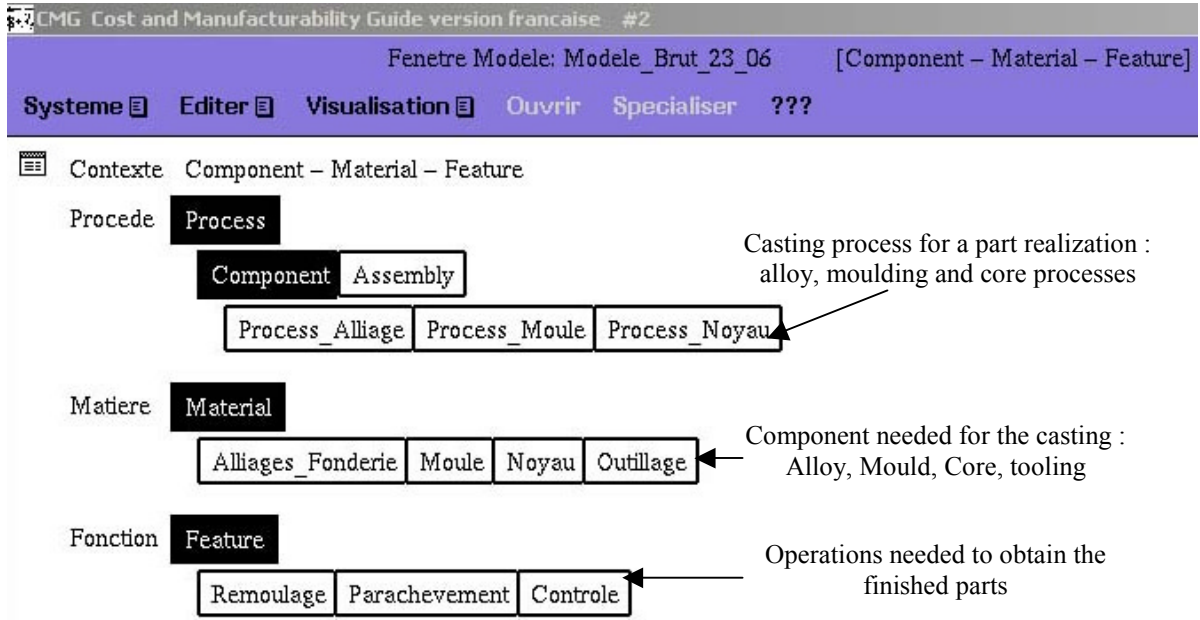

*Fig.2 : Sand casting Modelling*

## 3.2 Sand casting modelling

Based on this analysis we created a generic model using Cost Advantage Software. The first step closely defines the production process dedicated to this industry. The master parameters acting upon the product cost must be identified in order to enrich the cost semantic of the model.

The generic approach quickly highlights the problems of the contexts characteristics. How to define a significant cost for the part and with which level of detail in order to be generic? We will not answer this question, but we will present the paths or solutions we used.

First, let us describe the sand casting organisation through the part life cycle in the enterprise. This model is based on the SMC Colombier Fontaine foundry (France), from AFE Metal group. We will reduce our cost model to the primary parts through its life cycle in the enterprise and focus on the production phase, from the sand elaboration, the tooling machining and the parts perfecting.

## 3.3 Step induced by the use of Cost Advantage

Figure 3 presents a functional view of the process with the compound (raw material, tooling) and the elements needed to manufacture a part linked with the major indicators dealing with the final cost (loss, scrap ratio, production rate).

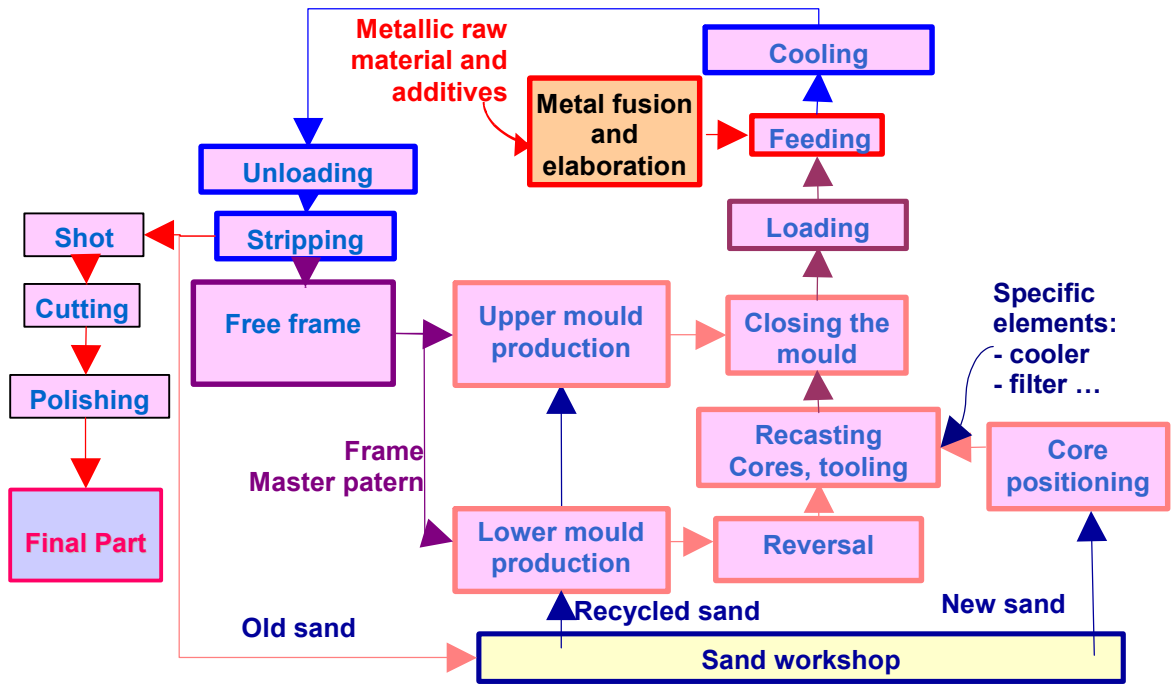

*Fig.3: Process view of a sand casting production cycle*

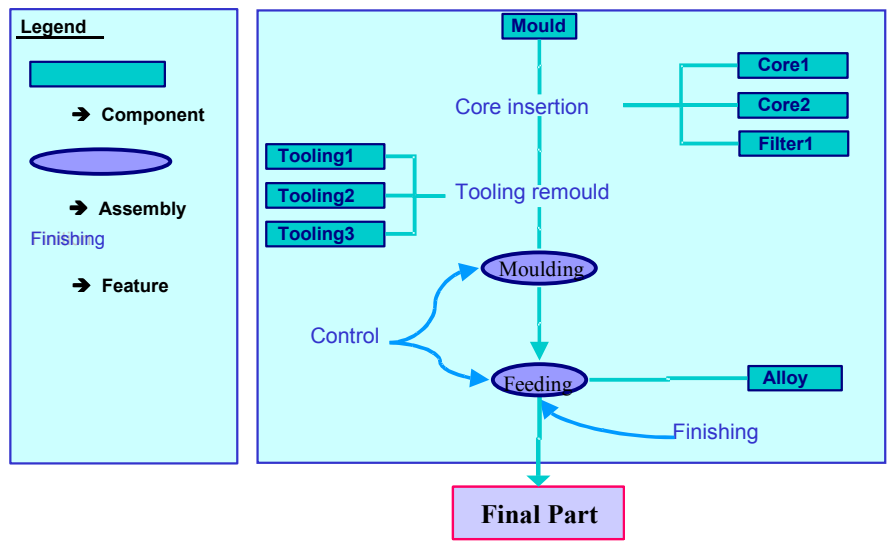

*Fig.4: Cost Advantage modelling example, at the assembly level*

Figure 4 represents a transposition under the concepts of Cost Advantage of this model gathering the three levels of entities defined in the software. To illustrate this, for example, the mould, the tooling and the cores are components required to carry out the assembly named moulding by the operation (feature) of remould. It is thus necessary to define the final part, to carry out the two assemblies, which are the moulding then the casting.

In term of model design, the functional view identifies the assemblies needed, it is then necessary to define the components and choose and define the related operations. An ascending step must be practiced, starting with the components up to the definition of the assemblies. The costs calculation is presented in figure 5 and the structure of the implemented data in figure 6.

Calculations are simply taking into account volumes of material, rates of production, losses and the machine and labour costs. To say nothing about the difficulty in knowing the exact parameters, the problems we met were model organization, more than process modelling.

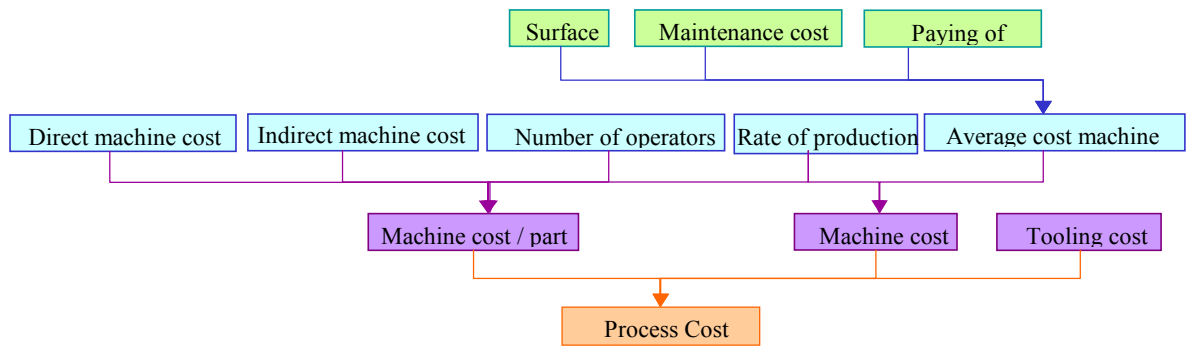

*Fig.5: Process cost structure sample*

The rules of calculation then implemented will make possible to the future user to inform only the relevant data about its study. Indeed, only the operational process, rates, dimensions, numbers of cores (etc.) will be required (or deduced directly in a CAD software) to allow an automatic calculation of the cost of the part according to its particular characteristics.

## 4 Discussion

The principal remark on this model, implemented with Cost Advantage platform, comes to that it doesn't integrate the global aspect of the costs management. For instance, the indirect share due to the development and the design of the tools (master pattern, cores boxes…) isn't taken in account.

But let us keep in mind the framework of use of such a tool: to help the designer to achieve a cost objective related to expected technical functions. He can then propose design modifications (joint section, cores, quality…) or process (number of parts in the mould…) to achieve its goal. Moreover, when the whole partners validate this design, the parameters costs are fixed (in agreement or not with the objective laid down initially), the cost is well known and will not be any more a consequence of later decisions.

During this work we have identified a principal difficulty, which for this modelling impact the multiplicity of the elements on the availability for their characterization and for their organization. Even if the manufacturing sand casting process seems simple, it uses many components (alloy, cores, mould…), and we limited the definition in term of model refinement since each one of these components could be the subject of a finer modelling. We tried to choose this limitation in order to represent the general process without going closer into enterprise specificities. For example, the cores or fusion process of realization are not completely defined since depending on the machines, uses and other specificities of the workshop. In fact this remark is shareable for all the components entering into the realization of the finished part. But we think that we defined a basic minimal skeleton, transposable from one company to another using the sand casting process.

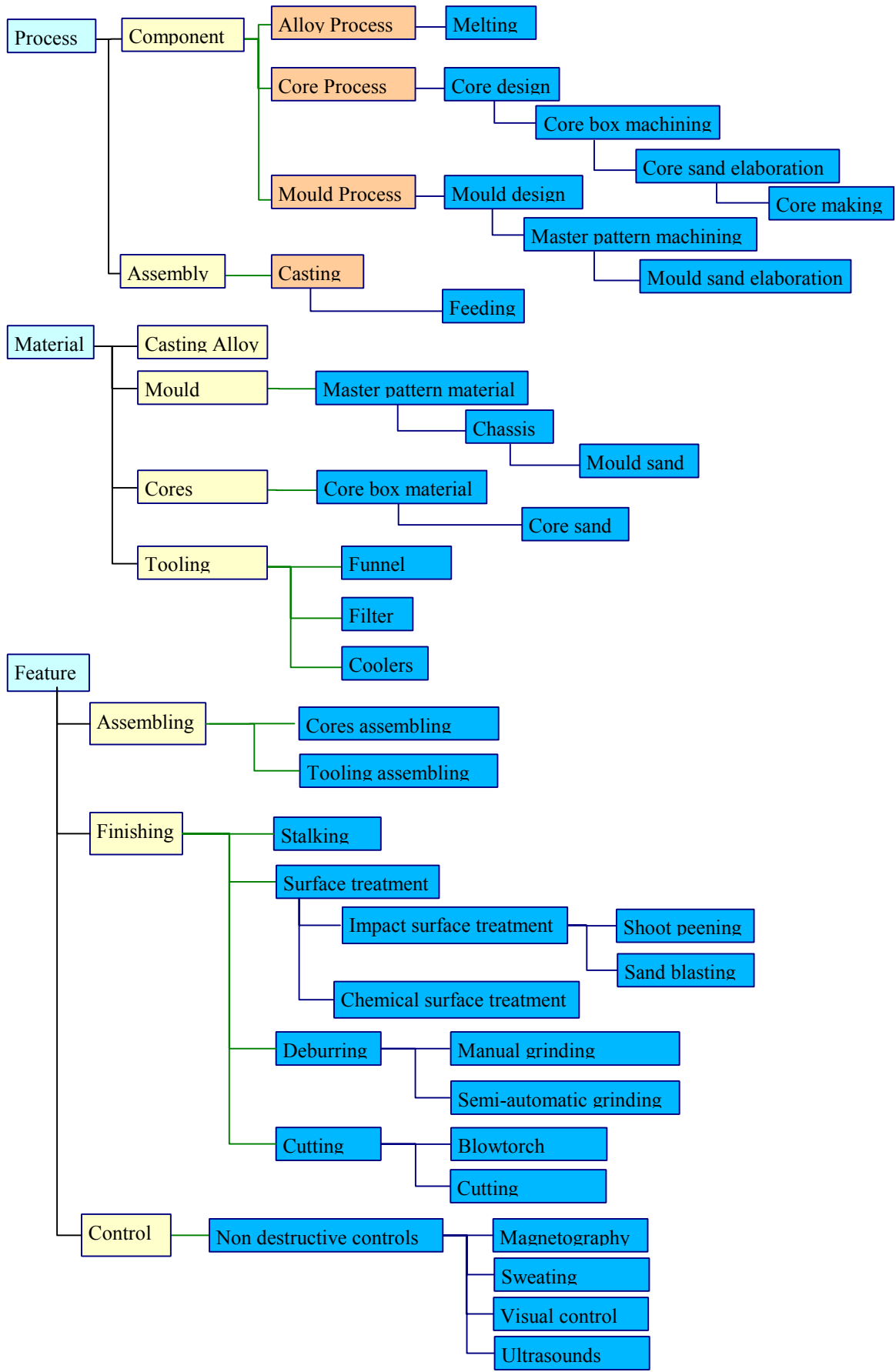

*Fig.6: Structure of data with Cost Advantage*

## 5  Conclusion

To conclude on this work, we started to apprehend logic of cost oriented modelling through a tool using the concept of cost entity. In order to ensure a generic aspect of our work, we deliberately limited the details of the operations, components and assemblies. Indeed, the development of these elements takes into account many parameters that it seemed to us initially overflowing to define.

We thus defined a structure that we think minimal, as well as indicators necessary to evaluate the whole costs without the indirect part. The two next steps of this study are in the first hand among other sand casting companies, to apply this modelling, configure the model with the existing processes and informing the exact values of the indicators. But also in the other hand, to calibrate the model and the results on real studies already done. These two last steps allow comparing the effectiveness of the various companies and could be used as Benchmark. A foreseeable difficulty is the possibility of reaching this information. Moreover, these factors are often managed in a total cost accounting. As a result, the efficient indicators may be sunk in a not very transparent accounting system or may be aggregated with not relevant others.

The other significant continuation to give to this work is the taking into account of the global costs mainly related to the indirect shares (structural). Our introduction puts forward the lack of management of these aspects and our first approach did not give place to a better control of these factors. However the work is done and the workers must be paid (designer, maintenance, buyers, logistics…) even if their work is not as well managed through a cost management system. A better specification (by the means of indicators, of metric) of the tools design phases, tools lifespan, etc. could integrate a real cost of the complete series. The question of the relevance of the tool used for this type of approach arises then. Some solutions come from the use of single or very limited number of cost inductors such as the time and they define a minimal global enterprise cost per hour to balance its financial objectives. Such an approach allow a multi-level management of the parts impact and give real time information to asses the enterprise objectives and manage strategies and operational decisions.